\documentclass[aps,pra, twocolumn,groupedaddress, 10pt]{revtex4-1}
\usepackage{graphicx}% Include figure files
\usepackage{dcolumn}% Align table columns on decimal point
\usepackage{bm}% bold math
\usepackage{color}
\usepackage{amsfonts,amsmath,amsthm,graphicx}
\usepackage{hyperref}

\usepackage{epstopdf}
\begin{document}

%\preprint{}

\title{Simultaneous accurate determination of both gravity and its vertical gradient}


\author{R. Caldani}
\author{K.X. Weng}
\thanks{Center for optics and optoelectronics research, College of Science, Zhejiang University of Technology}
\author{S. Merlet}
\author{F. Pereira Dos Santos}   
\email[]{franck.pereira@obspm.fr}
\homepage[]{https://syrte.obspm.fr/spip/}
%\thanks{}
\affiliation{LNE-SYRTE, Observatoire de Paris, Universit\'e PSL, CNRS, Sorbonne Universit\'e, 61 Avenue de l'Observatoire, 75014 Paris, France}


\date{\today}

\begin{abstract} 
We present a method for the accurate measurements of both the gravity acceleration and its vertical gradient using a dual atom interferometer, in principle free from any uncertainty related to the absolute or relative positions of the two atomic samples. The method relies on the use of a dual lock technique, which stirs simultaneously the chirp rate applied to the frequency difference between the interferometer lasers to compensate the gravity acceleration, and the frequency jump applied to the lasers at the mid pulse of the interferometer to compensate for the gravity gradient. This allows in the end to determine the two inertial quantities of interest in terms of frequencies.

\end{abstract}

%\keywords{}

\maketitle

\section{Introduction}

Atom interferometry techniques based on the manipulation of atomic wavepackets with light beamsplitters have led to the development of highly sensitive and accurate inertial sensors, accelerometers and gyrometers, whose performances compete favorably with, and in many cases surpass, state of the art conventional sensors~\cite{Peters2001, Gillot2014, Savoie2018, Karcher2018, Freier2016, Hu2013}. They find today applications in various fields, spanning from fundamental science and metrology to geophysics, exploration, monitoring, navigation and civil engineering \cite{Bouchendira2011, Jiang2012, Lautier2014, Zhou2015, Barrett2016, Thomas2017, Jaffe2017, Bidel2018, Parker2018, Geiger2018}.
The maturity of this technology has reached the level of industrial transfer and first commercial sensors are now available~\cite{Menoret2018}. Yet, the technology still has a large potential for improvement with the development of new techniques based on ultracold atoms~\cite{ Abend2016, Hardman2014, Karcher2018}, large momentum transfer beamsplitters~\cite{Clade2009, Muller2009, Chiow2011, Asenbaum2017, Mazzoni2015, Plotkin2018} and the long interrogation times available in large scale infrastructures~\cite{Canuel2018, Kovachy2015} or in space environment~\cite{Becker2018, Douch2018, QUEST}. 

Among these sensors, gradiometers have been developed, which measure gravity gradients out of the differential acceleration of two vertically~\cite{ McGuirk2002, Yu2006, Sorrentino2014} or horizontally~\cite{Biedermann2015} separated accelerometers. They have been applied in laboratories to the measurement of the gravity field induced by well characterized source masses, allowing for the determination of $G$ at the $10^{-4}$ level~\cite{Fixler2007, Rosi2014}. Moreover, their ability to reject common mode vibration noise also makes them particularly suitable for onboard gravity measurements, such as on ships, planes or satellites~\cite{Douch2018, Carraz2014}.

In these sensors, the relevant signal is extracted from the difference of the atom interferometer (AI) phases of the two accelerometers. Various methods have been developed for such an extraction out of eventually noisy individual acceleration measurements, such as ellipse-fitting methods~\cite{Foster2002}, Bayesian statistical analyses \cite{Stockton2007}, direct extraction of the differential phase~\cite{Bonnin2015}, the use of three simultaneous atom interferometers~\cite{Rosi2015} or the active differential phase extraction method of \cite{Chiow2016}. The correlation with acceleration measurements of auxiliary classical sensors~\cite{Caldani2017}, or the operation of the accelerometers in moderate~\cite{Sorrentino2012} or low~\cite{Duan2014} levels of vibration noise, allows in addition for the determination of both individual phases. Recently, a new method based on the compensation of the differential phase via a well-controlled frequency jump (FJ) applied to the interferometer lasers has been proposed~\cite{Roura2017} and demonstrated~\cite{Amico2017, Overstreet2018}, which allows for an accurate determination of the gravity gradient. By contrast with the other methods previously mentioned, this method does not require the precise knowledge of the gradiometer baseline, \textit{i.e.} the distance between the two accelerometers. 

Here we demonstrate a method based on the simultaneous determination of both the gravity acceleration and its vertical gradient in a dual gravi-gradiometer instrument. The method combines the dual lock method demonstrated in~\cite{Duan2014}, and the precise compensation of the differential phase phase using the FJ method~\cite{Biedermann2015}. It allows for the measurement of both quantities at their best level of stability, in particular thanks to the efficient rejection of the common mode vibration noise in the gradiometer measurement. The stability of the gravity acceleration measurements is limited by residual vibration noise, while the stability of the gravity gradient measurements is limited by detection noise. The two quantities of interest are finally derived out of frequency measurements, providing their absolute and SI traceable determinations.

\section{Experimental setup}\label{section2}

\begin{figure}[!h]
	\includegraphics[width=8.5cm]{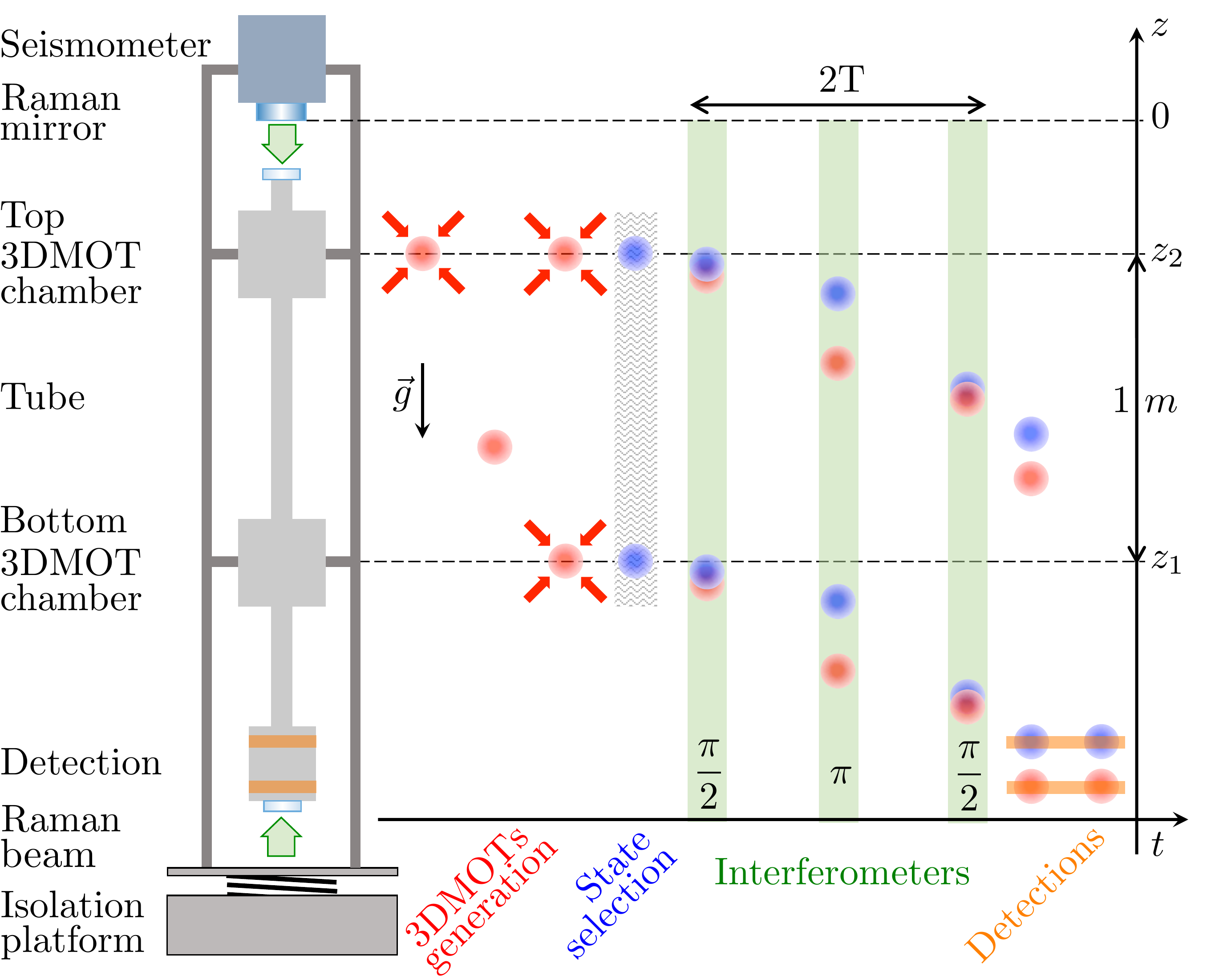}
	\caption{Scheme of the experimental setup and measurement sequence. Clouds displayed as blue circles are in the $\mid$F$= 1\big>$ state, clouds displayed as red circles in $\mid $F$=2\big>$. }
	\label{fig:gradio}
\end{figure}

The experimental setup and the time chart of the measurement sequence are displayed in figure~\ref{fig:gradio}. The vacuum chamber is composed of two vertically separated cold atoms preparation chambers, connected via a tube, allowing for the generation of two individual 3D magneto-optical traps (MOTs). We start by loading a first $^{87}$Rb atomic cloud in a 3D-MOT in the top chamber, out of the flux of a 2D-MOT (not represented on figure~\ref{fig:gradio}). We collect about $10^7$~atoms in 480~ms, cool them down in a far detuned molasses down to 2~$\mu$K, before releasing them in free fall. After about 450~ms of free fall, the first cloud is recaptured in the bottom chamber MOT and we start loading a second cloud in the top chamber MOT for 200 ms. Both clouds are finally simultaneously cooled down to 2~$\mu$K and released in free fall. At the end of this preparation phase, which lasts about 1.2~s, we end up with a few $10^6$ atoms in each cloud in the $\mathrm{|F=2\big>}$ hyperfine ground state. The atoms are then selected in the $\mathrm{|F=1, m_F=0\big>}$ magnetic state using a combination of microwave and pusher pulses. During their free fall, Mach-Zenhder interferometers are performed using a sequence of three counterpropagative Raman laser pulses ($\pi/2$-$\pi$-$\pi/2$)~\cite{Kasevich1991}, separated by free evolution times $T=80$~ms. Finally, the whole sequence ends at the bottom of the experiment with the successive time of flight fluorescence detection of the populations of the two interferometers output ports , thanks to the state labeling method~\cite{Borde1989}.

\begin{figure}[!h]
	\includegraphics[width=8.5cm]{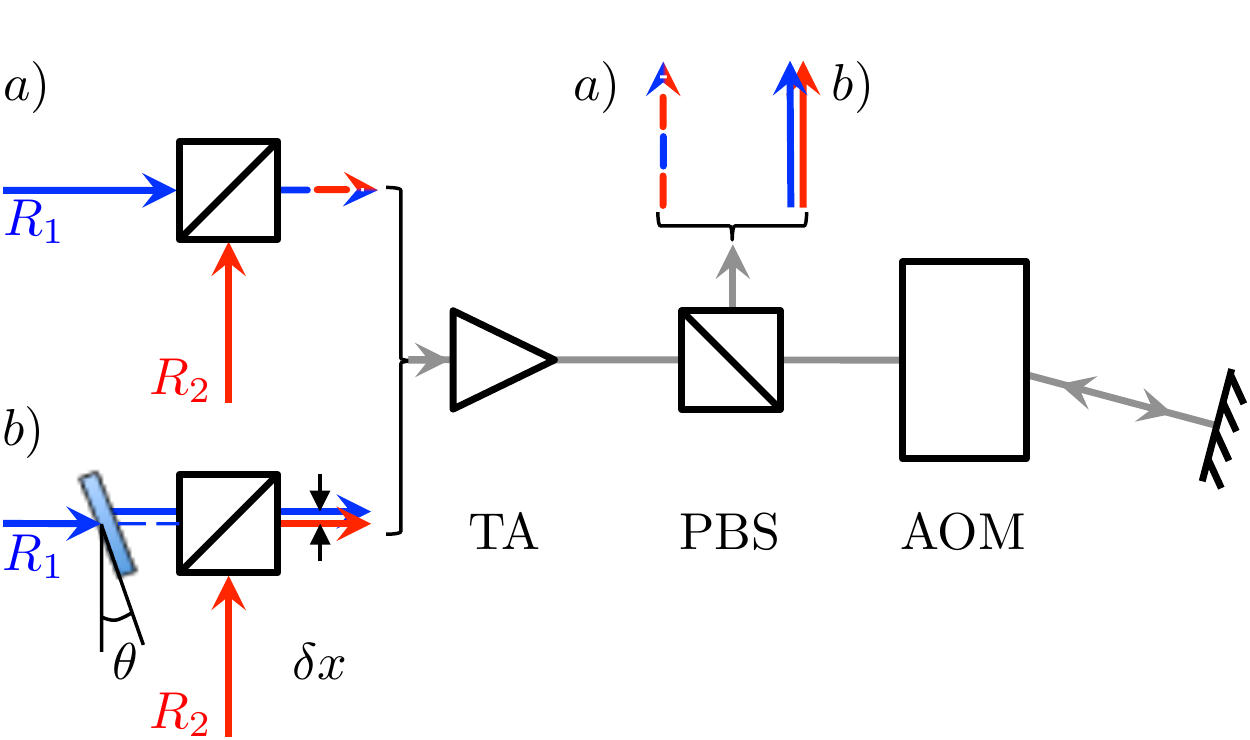}
	\caption{Raman beams ($R_1$, $R_2$) optical setup. a) Perfect superposition. b) Imperfect superposition induced by an additional glass plate. PBS: polarization beam splitter, TA: tapered amplifier, AOM: acousto optic modulator.}
	\label{fig:AOM_window}
\end{figure}

The two Raman lasers are injected from the bottom of the experiment and reflected on a mirror attached to a seismometer placed on top of the experiment. This mirror defines the position of the effective wavefront of the Raman lasers which constitutes the reference for the acceleration measurement. To reduce the mirror position noise, the whole experiment is placed on a passive isolation platform. In addition, its motion is recorded by the seismometer, which allows to correct the interferometer measurements from residual vibrations, thus improving the measurement stability~\cite{Gouet2008}. In this retroreflected lasers configuration, four beams are actually send onto the atoms. For properly adjusted lasers polarizations, one can select counterpropagating transitions with Raman wavevectors pointing either upwards ($k_\uparrow$) or downwards ($k_\downarrow$), depending on the sign of the Doppler shift applied to the frequency difference between the Raman lasers~\cite{Louchet2011}. 

Two different optical systems are used in our experiment. A commercial bench based on frequency doubled telecom lasers~\cite{muquans} supplies the 2D and upper 3D-MOT while a homemade bench based on semi-conductor lasers, supplies the bottom 3D-MOT, the Raman and detection beams~\cite{Merlet2014}. The Raman beams are simply generated out of the lasers used for laser cooling using a double pass AOM, such as displayed in figure~\ref{fig:AOM_window}.

\section{Atomic interferometers phase control}

We start by recalling the expression of the interferometer phase for a gravimeter.
\begin{equation}
\Phi=\phi(0)-2\phi(T)+\phi(2T)=kgT^2
\label{eqPhase}
\end{equation}
where $\phi(t)$ is the Raman laser phase difference at time $t$ at the center of the atomic wavepacket, $k$ is the effective Raman wavevector and $T$ the time separation between pulses.

In practice, a frequency chirp (FC) is applied on the Raman frequency difference in order to compensate for the Doppler shift and keep the resonance condition satisfied for the Raman transitions~\cite{Louchet2011}. This adds to the AI phase a contribution $aT^2$, where $a$ is the angular frequency chirp, which exactly compensate the gravity phase shift for $a=-kg$. This equation provides an accurate determination of $g$ as it relies on frequency measurements only. Our gradiometer is obtained by simultaneously interrogating with the same Raman lasers two atom sources separated by a baseline $L$. The gravity gradient can be obtained from the differential phase between the two AIs.
\begin{equation}
\Delta\Phi=\Phi_2-\Phi_1=kg_2T^2-kg_1T^2=k\gamma LT^2
\label{Dphase}
\end{equation}
where $\Phi_i$ and $g_i$ are the phase and the gravity acceleration for the AI $i$ ($i$=1 for the bottom and 2 for the top AI) and $\gamma$ the vertical gravity gradient. 
The phase difference between the two AIs can be modified in a controlled and accurate way by performing a frequency jump (FJ) onto one of the Raman pulses~\cite{Biedermann2015}. The resulting change in effective wavevector $\delta k$, when properly adjusted, can even compensate for the effect of the gravity gradient \cite{Roura2017}. 
Applying a frequency jump (FJ) $\Delta\nu$ on the AIs $\pi$ pulse results in additional phases shifts $\Delta\Phi_i^\mathrm{FJ}$, proportional to $\Delta\nu$ but different for the two AIs, given by:
\begin{equation}
\Delta\Phi_i^\mathrm{FJ}=K_i\Delta\nu
\label{eqdphi}
\end{equation}
with $K_i=8\pi z_i/c$, where $z_i$ is the atoms distance to the mirror at the $\pi$ pulse and $c$ the velocity of light.
In the presence of both a FC and FJ, the AIs phase of equation~\ref{eqPhase} can be written as: 
\begin{equation}
\Phi_i=kg_i T^2+aT^2+K_i\Delta\nu
\label{eqPhase2}
\end{equation}
Using the two independent control parameters $a$ and $\Delta\nu$, one can set both AIs phase simultaneously to 0. This corresponds to the control values:
\begin{equation}
\Delta\nu=\frac{k(g_1-g_2)T^2}{K_2-K_1}
\label{Dnu}
\end{equation}
and
\begin{equation}
a_s=-k\bigg[\frac{K_2g_1-K_1g_2}{K_2-K_1}\bigg]=-kg_s
\label{achirp}
\end{equation}
where $g_s$ (resp. $a_s$) is a synthetic $g$ (resp. $a$) value resulting from a linear combination of the gravity accelerations of the two clouds. Assuming that the gravity difference betwen the two AIs depends only on the gravity gradient, we can write $g_i=g_0+\gamma z_i$ with $g_0$ the gravity acceleration at the mirror position. Considering the scaling of $K_i$ with $z_i$, the two previous equations lead to:
\begin{equation}
\begin{array}{ll}
\Delta\nu_\gamma=-\gamma \dfrac{kT^2c}{8\pi}\\
g_s=g_0\\
\end{array}
\label{eqadgcomb}
\end{equation}
with $\Delta\nu_\gamma$ the frequency jump that compensates for $\gamma$. This method thus provides accurate determinations of both the gravity acceleration (at the mirror position) and the gravity gradient, independent from the baseline, and from the positions of the two sources.

\section{Dual numerical integrator implementation}

We now present how we implement the dual lock on the experiment in order to perform a simultaneous measurement of $g_s$ and $\gamma$. With equation~\ref{eqPhase2} we express the transition probabilities we derive from the populations measurements in the two output ports:
\begin{equation}
P_i=A_i+\frac{C_i}{2}\cos\Big(kg_iT^2+aT^2+\frac{8\pi \Delta\nu z_i}{c} \Big)
\label{eqProba1}
\end{equation} 
where $A_i$ is the offset and $C_i$ the contrast of the i$^{th}$ interferometer. For $g_i=g_0+\gamma z_i$, the phase of the previous equation can be factorized as:
\begin{equation}
P_i=A_i+\frac{C_i}{2}\cos\Big(\delta aT^2+\delta(\Delta\nu) \frac{z_i}{c}\Big)
\label{eqProba2}
\end{equation} 
where $\delta a=kg_0+a$ and $\delta(\Delta\nu)=8\pi\Delta\nu+k\gamma cT^2$ are respectively the errors on the determination of the FC and FJ. To lock those parameters to 0, we modulate the phase of the AIs by alternatively adding offset phases of $\pm \pi/2$, so as to operate both AIs at mid fringe where their sensitivity is optimal, as performed in the conventional gravimeter measurements~\cite{Merlet2009} or in atomic fountains. For small phase errors, the difference between the transition probabilities measured at mid fringe on the left and on the right of the central fringe then gives:
\begin{equation}
P_{i-}-P_{i+}=C_i\Big(\delta aT^2+\delta(\Delta\nu) \frac{z_i}{c}\Big)
\label{eqProba3}
\end{equation}
Combining those differences for the two AIs allows to determine independently the errors $\delta(\Delta\nu)$ and $\delta a$:
\begin{equation}
\begin{array}{ll}
\delta(\Delta\nu)=\dfrac{c}{z_1-z_2}\Big(\dfrac{P_{1-}-P_{1+}}{C_1}-\dfrac{P_{2-}-P_{2+}}{C_2}\Big) \\
\delta a=\dfrac{1}{(z_2-z_1)T^2}\Big(\dfrac{P_{1-}-P_{1+}}{C_1}z_2-\dfrac{P_{2-}-P_{2+}}{C_2}z_1\Big) \\
\end{array}
\label{eqErr}
\end{equation}

These are used as error signals in an integrator digital locking loop: at each measurement cycle, the values of $a$ and $\Delta\nu$ are corrected by $\pm G_a\delta a$ and $\pm G_{\Delta\nu} \delta(\Delta\nu)$, $G_a$ and $G_{\Delta\nu}$ being the gains of the integrator loops. The resulting dual lock makes $a$ and $\Delta\nu$ converge towards $a_s$ and $\Delta\nu_\gamma$.

Note that in this lock method the determination of the frequency jump error signal is based on the direct extraction of the differential phase. The efficient rejection of common mode vibration noise thus requires the accurate knowledge of the interferometers contrasts, whose slow fluctuations can for instance be tracked via their periodic monitoring. 

\section{Results}

The FJ on the Raman lasers is applied by changing the frequency of the AOM, which is used, in addition to offset the lasers detuning and to switch on and off the Raman pulses. Its frequency is set by a DDS, which ensures precise numerical control and fine tunability. With our AIs separation of 1~m, a differential phase of $\Delta\Phi=\pm1$~mrad corresponds to a FJ of $\pm12~$kHz.
\begin{figure}[!h]
	\includegraphics[width=9cm]{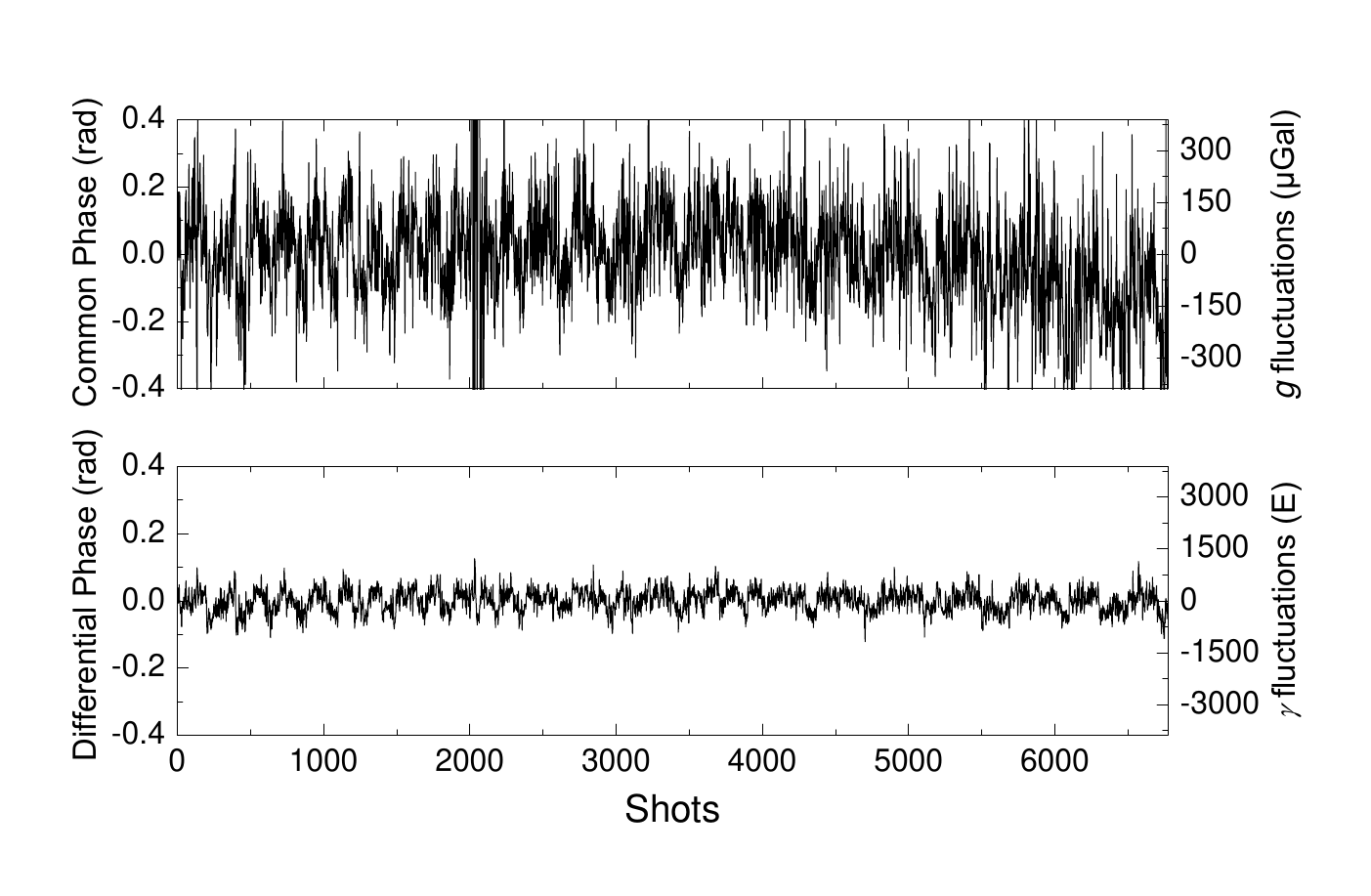}
	\caption{Results of the dual lock measurement. Top: Common phase and corresponding gravity acceleration measurement. Bottom: Differential phase and corresponding gravity gradient measurement.}
	\label{fig:duallock}
\end{figure}
We performed a dual lock measurement for interferometer duration $2T=160$~ms, following the measurement sequence described in section~\ref{section2} with a cycle time of $1.8$~s. Figure~\ref{fig:duallock} shows the fluctuations of the common ($a_s T^2$) and differential ($8\pi L \Delta\nu / c$) phases, in units of rad, and their corresponding gravity acceleration ($g_s$) and gravity gradient ($\gamma$) fluctuations in units of $\mu$Gal and E (1~$\mu$Gal~$=10^{-8}$m/s$^2$ and 1~E $=0.1~\mu$Gal/m~$=10^{-9}$/s$^2$).
The clear difference in the amplitude of these fluctuations arise, as expected, from the suppression of common mode vibration noise in the differential measurements. We calculate Allan standard deviations of 190 and 91~$\mathrm{mrad/\sqrt{Hz}}$ for the common and differential phases, corresponding to gravity acceleration and gravity gradient stabilities of 180~$\mathrm{\mu Gal/\sqrt{Hz}}$ and 890~$\mathrm{E/\sqrt{Hz}}$ respectively.

We then compared the results of the dual lock measurement with the results of individual gravimeter measurements performed at the top ($g_2$) and at the bottom ($g_1$), with the same interferometer duration $2T=160$~ms. From these two measurements, we deduce the gradient $(g_2-g_1)/L$ and via a linear extrapolation $g_0=g_i-z_i(g_2-g_1)/L$ and compare these determinations with the values $\gamma$ and $g_s$ measured with the dual lock. Table~I presents the results of the measurements of the gravity accelerations and gradient obtained with double lock and single locks methods. Average gravity values are given with respect to $g_1$, which we take here as a reference. While the two determinations of the gravity gradients are in agreement, the value $g_s$ is about 100~$\mu$Gal away from $g_0$. Note that all the above mentioned measurements are in fact performed by interlacing different measurements configuration, and averaged over the two opposite $k_{\uparrow}$ and $k_{\downarrow}$ wavevector directions, which rejects a number of systematic effects and drifts~\cite{Louchet2011}. 
\begin{table}
\begin{tabular}{ c c || c c || c c }
 \multicolumn{2}{c}{Dual Lock}&\multicolumn{2}{c}{Single Locks} & & \\
 \hline
 $g_s$ & $\gamma$ & $g_1$ & $g_2$ & $g_0$ & $g_s^{\text{calc}}$ \\
 ($\mu$Gal) &  ($\mu$Gal/m) & ($\mu$Gal) &  ($\mu$Gal) & ($\mu$Gal) &  ($\mu$Gal) \\
 \hline
 -524(4) & -273(10) & 0(0) & -267(4) & -435(7) & -509(12) \\
\hline
\end{tabular}
\caption{Comparison between the results obtained with the dual lock (FC and FJ) and single locks (FC) methods. Values are given relatively to $g_1$ which we use here as a reference.}
\end{table}

\begin{figure}[!h]
	\includegraphics[width=9cm]{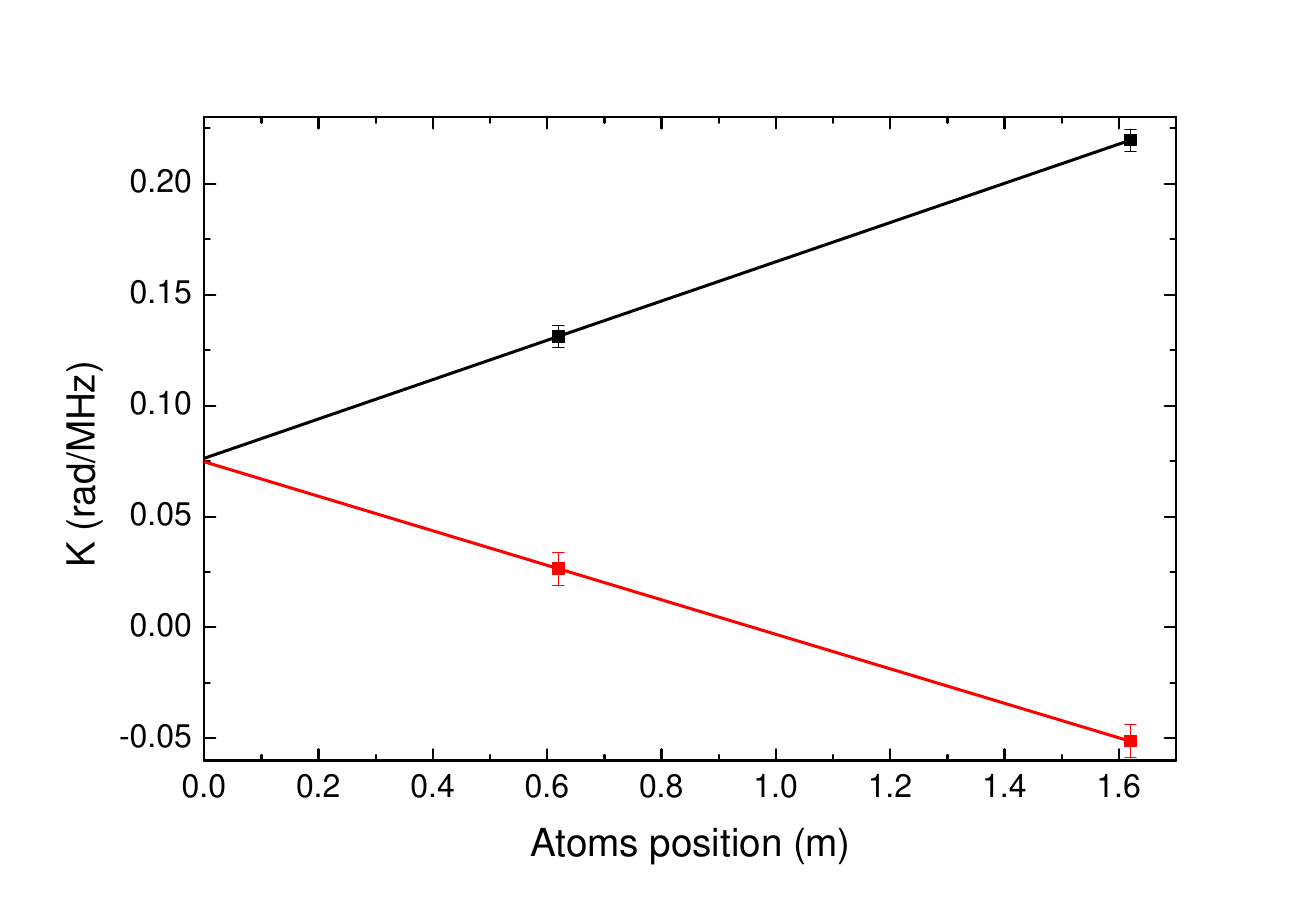}
	\caption{$K$ parameters as function of the distance to the mirror. The slopes are ($0.088(8)$~rad/MHz/m) for $k_{\uparrow}$ (black) and ($-0.078(10)$~rad/MHz/m) for $k_{\downarrow}$ (red).}
	\label{fig:K}
\end{figure}

Given equation~\ref{achirp}, a possible explanation for this difference might be that the $K$ coefficients differ from their expected values. We then performed measurements of the change of the interferometers phase $\Delta\Phi^\mathrm{FJ}$, individually for each clouds, as a function of the FJ. $\Delta\Phi^\mathrm{FJ}(z)$ was found to scale linearly with $\Delta\nu$ as expected. Results of the slopes, corresponding to the $K_i$ coefficients, measured for the two AIs and for the two $k_{\uparrow}$ and $k_{\downarrow}$ Raman wavevectors directions, are displayed on figure~\ref{fig:K}. We obtain differences between $K_1$ and $K_2$ coefficients of 0.088(8) and -0.078(10)~rad/MHz, depending on the direction of the effective wavevector, in agreement with the expected value of $\pm 8\pi L/c$. But, the linear extrapolation of the coefficients at the mirror position results in values of about 0.075(4) rad/MHz, different from the expected values (null). Using these measured values for $K_1$ and $K_2$, we calculated the expected value for $g_s$  in table~I, which we finally found in agreement with the measured one.

\section{Total phase shift}

\begin{figure}[h]
	\includegraphics[width=9cm]{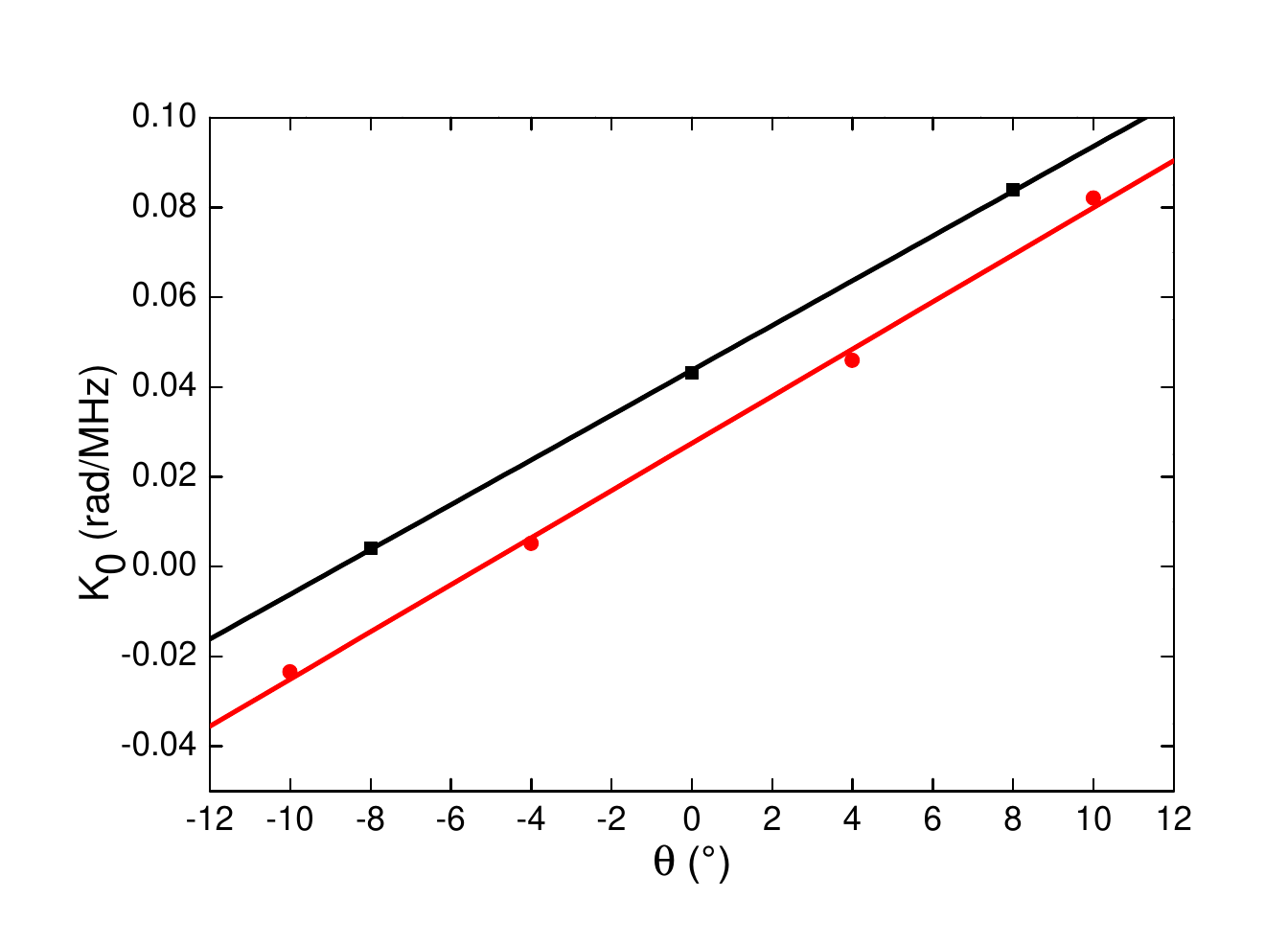}
	\caption{Extrapolated offset value $K_0$ as a function of the glass plate angle $\theta$. The two sets of measurements, represented as circles and squares, are taken at a 3-days time separation. Lines are linear fits to the data.}
	\label{fig:K0}
\end{figure}

We attribute this phase shift to an imperfect overlap between the centers of the two Raman beams in the AOM. A shift $\delta x$ in position of the beams inside the crystal, and along the direction of propagation of the acoustic wave, leads to a phase difference between the two Raman beams after the double pass of $\Delta \phi=2\times 2\pi \nu_\mathrm{AOM}\delta x /v_s$, where $v_s$ is the velocity of the acoustic wave in the crystal. A change $\Delta\nu_\mathrm{AOM}$ at the $\pi$ Raman pulse on the AOM thus results in a phase difference on the interferometer phase of: 
\begin{equation}
\delta(\Delta\phi)=2\times 4\pi\frac{\Delta\nu_\mathrm{AOM}}{v_s}\delta x
\label{eqAOM}
\end{equation}
where the extra 2 factor is due to the impact of the Raman phase difference at the $\pi$ pulse in the interferometer phase equation~\ref{eqPhase}. We find $\delta(\Delta\phi)/\Delta\nu_\mathrm{AOM}=75$~mrad/MHz, which, for $v_s=4200$~m/s, corresponds to a shift of $12.5~\mu$m only. Such a shift is possible since the AOM is placed after a free space tapered amplifier which is not a good spatial filter.
To highlight this effect, we deliberately modified the overlap between the beams by placing a glass plate on the path of one of the Raman beam before the superposition with the second (Fig~\ref{fig:AOM_window} b). We then measured again (for $k_\uparrow$ only) the $K$ parameters for several orientations $\theta$ of the glass plate to control the Raman beams overlap, and extracted the corresponding $K_0$ offset value at the mirror position. 
The results displayed in figure~\ref{fig:K0} shows that varying the glass plate angle $\theta$ modifies the offset value. One can in principle compensate for the superposition mismatch by carefully adjusting the plate angle, so as to nullify this offset. Unfortunately, this effect is found to vary over the long term, as evidenced by the different offsets in measurements obtained after a 3-days time interval. We attribute this to fluctuations of the optical bench alignment over time: the difference between the two curves on the figure corresponds to a drift in $\delta x$ as small as 3~$\mu$m in 3 days. 

\section{Conclusion and Discussion}
We have demonstrated a method for the simultaneous and accurate determination of the gravity acceleration and its vertical gradient independently from the baseline, thanks to controlled frequency chirps and frequency jumps applied to the Raman lasers frequencies. We have put into evidence the effect of the imperfect overlap between the Raman lasers, when using an AOM to stir the differential phase with changes of its driving frequency. This modifies the values of the stirring coefficients, but it does not compromise the accuracy of the measurements provided these coefficients are well determined.

As an alternative, the change in the Raman frequencies could be realized by changing the setpoint of the frequency lock of the master Raman laser. In our laser system, this could be done by applying a voltage offset onto the error signal of the lock loop. In that case, though, a calibration of the induced frequency change is required and the benefit of tying the inertial measurements to frequency measurements only is lost.

In this proof of principle demonstration of the proposed method, the stabilities obtained for the gravity acceleration and gravity gradients lie one to two orders of magnitude above state of the art performances. This is due to the combination of relatively large levels of vibration and detection noise, a long cycle time and a relatively short interfereometer duration. These performances will be improved by loading each chamber from independent 2D MOTs and by launching the atoms upwards. This will reduce the cycle time and allow increasing the interferometer duration up to 500 ms in our setup.

Finally, though demonstrated here for a relatively low level of vibration noise, the method can be adapted to much larger levels of vibration noise, even beyond the linear range of operation of the interferometers~\cite{Merlet2009}, by adequately exploiting the correlation between the interferometers and the sismometer such as in~\cite{Caldani2017}.

\section*{ACKNOWLEDGEMENTS}
This work was supported by CNES, DGA (Gradiom project),  the “Domaine d'Int\'er\^et Majeur” NanoK of the R\'egion Ile-de-France, the CNRS program “Gravitation, R\'ef\'erences, Astronomie, M\'etrologie” (PN-GRAM) co-funded by CNES, and by the federative action GPhys of Paris Observatory. We thank ESA for the use of the laser system developed by Muquans in the frame of the ITT AO/1-8417/15/NL/MP. K.X.W. thanks the National Key Research and Development Programs of China (2016YFF0200206, 2017YFC0601602) and R.C. thanks the support from LABEX Cluster of Excellence FIRST-TF(ANR-10-LABX-48-01), within the Program “Investissements d'Avenir” operated by the French National Research Agency (ANR).

\end{document}